# Stacking Fault in Non-Close Packed System- Role of Interstitials at Pentahedron Voids in WC Simple Hexagonal

Alphy George*, T. Sreepriya, Arun Kumar Panda, R. Mythili, Arup Dasgupta and R. Divakar

*Metallurgy and Materials Group, IGCAR, HBNI Kalpakkam-603102*
*alphy@igcar.gov.in

**Abstract**

*Atomistic origin of stacking faults in non-close packed systems is a fundamentally distinct mechanism from the well-known close packed structures with ABC stacking, and represents an uncharted territory in material research. According to experimental data, stacking faults in simple hexagonal WC happen in {1-100} planes that are packed rectangularly and have ABAB stacking. This work identified the type of the defect and crystallographic behaviour by creating energetically relaxed potential atomistic models of stacking faults in WC. Experimental evidence supporting the rotation axis along stacking fault caused by variation in carbon ordering at the interstitial site has been established, in accordance with the theoretical model.*

Planar defects within a crystal caused by slips in stacking sequence of planes are known as stacking faults. Both the packing and the stacking of the planes determine the nature of stacking fault. The widely recognized stacking defect occurs in hexagonally packed planes with ABC stacking of closely packed fcc crystals (e.g., $\{111\}_{fcc}$), which results in an fcc to hcp transition at the faulty plane. As a result of the fault, the hexagonally packed planes piled as ABAB, the basal planes of the hcp structure along the $<0001>_{hcp}$ direction ($//\{111\}_{fcc}$). These occur frequently in super alloys, austenitic steels, and other fcc systems [1]. In NaCl-type transition metal carbides such as NbC [2], SiC [3], etc., where the carbon occupies the octahedral voids, there is yet another fcc to hcp transition at the fault. Preliminary classification of stacking faults in close packed system of fcc are intrinsic (ABCBCABC), extrinsic (ABCBABCABC), deformation fault (ABCABCABC - ABCBCABCA) and growth fault (ABCABCACBACB) [4]. Metal carbides of the M23C6 type with a fcc structure also show stacking defects in {111} planes [5].

Conversely, basal plane stacking faults cause a hcp to fcc transition and are prevalent in semiconductor materials such as GaAs, GaSb, InAs, Mg-Y alloy [6], ZnO, and CdS [7], as well as post-transition metal nitrides like GaN, InN, and AlN [8-10]. The extra plane provides multiple sub-arrangements for the fault, including ABABABCABAB, ABABABCBCBC, and ABABABCACAC.

In contrast to the {111} ⇔ {0001} transition in close-packed structures, stacking faults in non-close-packed systems exhibit diversities in the formation plane and their mechanism according to its crystal symmetry. The {001} stacking fault in $MoSi_2$ BCC Tetragonal (ABABAB- ABBABAB) [11], $Ni_3Ti$ [12], $Fe_2B$ [13], orthorhombic $FeSi_2$ [14], $BiI_3$ (Biotite) [15] are some of the examples. Stacking faults are the most common type of planar defect found in oxides or carbides with layered structures comprising two or more metal atoms [16]. Because of the additional D plane, some transition metal di-silicide structures, such as C54 with ABCD stacking, are more likely to develop multiple ways stacking faults [17]. A more advanced kinematical model with high resolution imaging has been proposed in the $(Fe,Cr)_7C_3$ system with P31c symmetry, suggesting that the defect is not in a single layer but rather in the stacking of atomic clusters [18].

The tungsten carbide (WC) stacking defect is distinct from the complex layered structures mentioned above as it is a simple hexagonal structure (P-6m2) with c/a ratio less than one (Figure 1a-b). Unlike hcp structure, which has ABAB stacking along [0001], hexagonally packed (0001)$_{WC}$ planes have AAA stacking along [0001]. Another important plane is the {1100}$_{WC}$ planes, which have ABAB stacking and rectangular packing. Because the tungsten atom planes are stacked one on top of the other along [0001] to form pentahedral (five faces) voids in the shape of triangular prisms with trigonal-prismatic interstices, the WC structure lacks tetrahedral or octahedral voids. Six anion atoms contribute to the creation of pentahedra: the first nearest neighbour (1nn) in the c-axis (2.83660 Å) and the second nearest neighbour (2nn) in the a-axis (2.90650 Å). Carbon was found in one of the two pentahedra voids in the hexagonal unit cell of WC, resulting in an anion coordination of six. Carbon position is exactly middle of trigonal-prismatic interstices gives another simple hexagonal unit for carbon lattice originated at *a* position. Another way of the structure interpretation is 'intercalated sub lattices of simple hexagonal of tungsten and carbon atoms' (Figure 1b).

Transmission electron microscopy results (Figure 1c) shows the planar defects in WC inherently occur in {1100} planes. Phase contrast image in [0001] zone axis (Figure 1d) proves all three sets of {1100} planes are equally feasible to have stacking faults. In contrast, the stacking fault in closed-packed hexagonal systems is restricted to the basal plane and absent from the prismatic plane. It must be emphasized that in simple hexagonal systems with c/a ratio < 1, the prismatic plane with rectangular packing is the densest plane, in contrast to hcp systems. The mechanism by which a stacking fault can be caused by the atomic arrangement of ABAB stacking on a rectangularly packed {1100} plane is not yet understood. The current work makes an effort to pinpoint the defect mechanism in this particular packing/stacking.

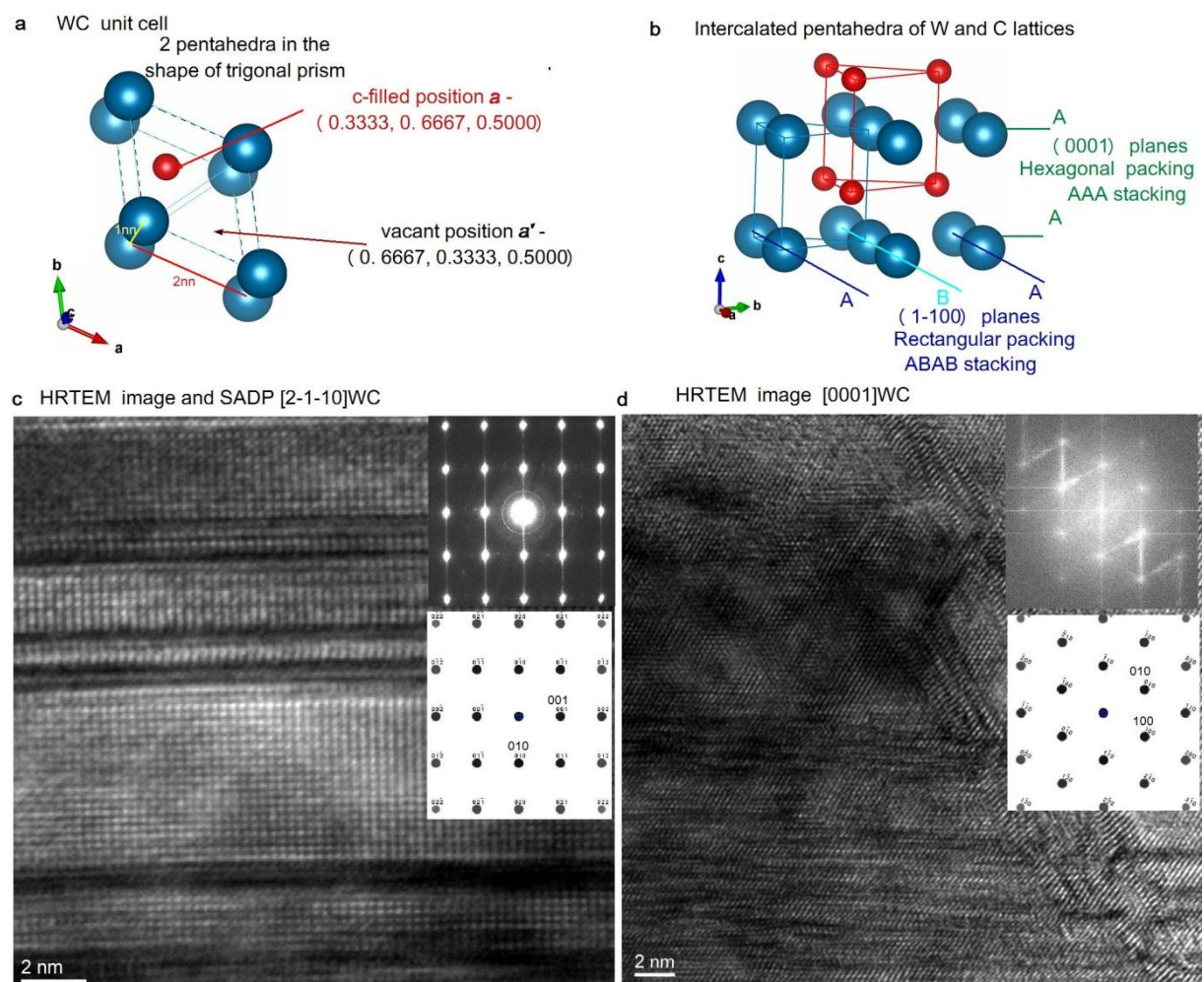

*Figure 1. (a-b) crystallographic features of WC simple hexagonal structure and (c-d) experimental evidence of stacking fault in {1100} planes of WC.*

**Potential defect models of WC crystal**

The problem has been approached by atomistic modeling of possible planar defects in {1100}$_{WC}$ crystal. WC unit cell has carbon filling in one of the pentahedra, say, position *a* (0.333, 0.667, 0.5) (Figure 1a). Other pentahedron is positioned at *a*': 0.667, 0.333, 0.5. These models postulate that fault formation could result from the planar reorganization of the carbon filling in positions *a* and *a*'. Hence the models are predicted from {1100}$_{WC}$ stacking of A*a*B*b*A*a*B*b*A*a*B to Model I: A*a*B*b*AB*b*A*a*B, Model II: A*a*B*b*A*aa*'B*b*A*a*B, Model III: A*a*B*b*A*a*'B*b*'A*a*'B, Model IV: A*a*B*b*AB*b*'A*a*'B, and Model V: A*a*B*b*A*aa*'B*b*'A*a*'B (Figure 2a and b). Model I and II, which show the growth defect from the faulted boundary with excess or deficient carbon atoms at the interstitials, Model III, which shows the inversion filling of carbon from *a* to *a*', and Model IV and V, which combine inversion filling at the fault with excess or deficient carbon atoms, form the basis of defect models.

Then, an energetically relaxed structure for the atomistic models generated above is theoretically calculated using first principle DFT codes. Computational screening techniques aim to detect the deviating atomic layers of the faulty plane. Figure 2c displays the relaxed structures of every model along [0001] zone axis. Because of the modification in the carbon filling, the completely relaxed structure of each model deviates from both the initial configuration and the original WC structure. Specifically, a compression on the faulty line is observed in models I and IV, where there is no filling of carbon atoms between the tungsten plane. Divergence is moderate for the inversely filled boundary shown in Model III. The defect plane of carbon at inversely filled interstitials causes a compression along 2nn (2.90650 -> 2.89864 Å) and a minor elongation along the c-axis. The structure (Model II and V) with carbon atom planes that are doubly filled in between tungsten planes relaxed to have the greatest displacement in the tungsten plane at the fault. At the faulted plane, both the spatial and angular coordinates change (Model II: a= 3.0062 Å, b= 2.9478 Å, and c= 2.9173 Å & α=β= 90º and γ= 86.22º; and model V: a= 3.0173 Å, b= 2.9216 Å, and c= 2.94974 Å & α=β= 90º and γ= 86.46º) The hexagonal packing also varies with double-filled carbon planes. Atomic projections shows a deviation from hexagonal symmetry to two-fold symmetry. Simulation shows that the defective plane should move the least along the c-axis. The misplaced tungsten plane and, by extension, the stacking defect are mostly caused by carbon filling in the pentahedral interstitials, as established by simulation analysis.

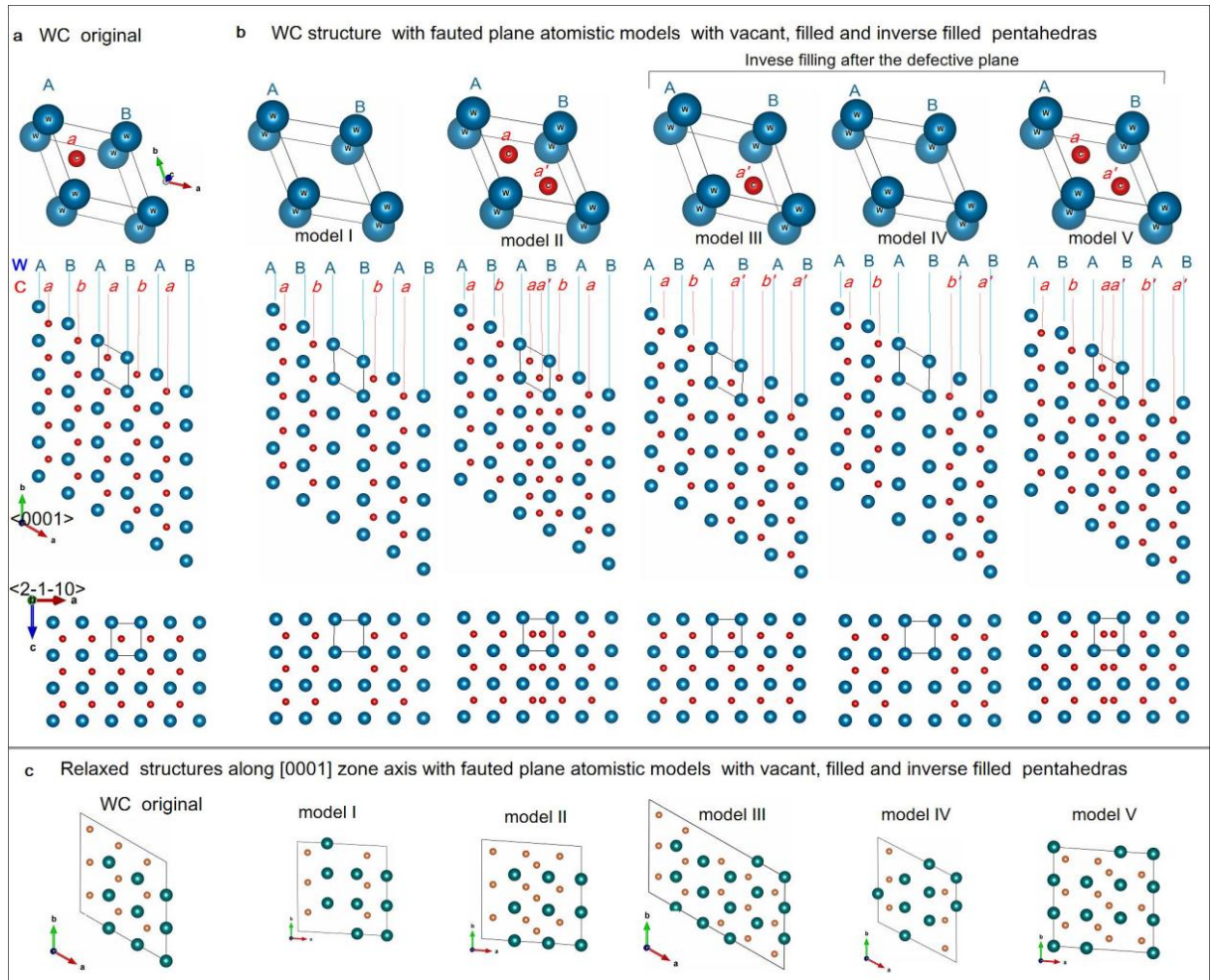

*Figure 2: Atomistic models created on WC parent structure based on carbon stacking along the 1100 plane (a) The original WC structure; (b) Five distinct models of the WC structure with defective plane by changing carbon filling. A and B indicate the tungsten position, and a, b, a', and b' are favorable interstitial carbon positions. Atomic projections of W and C in <0001> and <2-1-10> direction are shown for original and defected structures; and (c) energetically relaxed structures from all models by first principle calculations demonstrate atomic plane rearrangement as a result of flawed carbon filling.*

**Experimental confirmation of rotation axis at SF**

Comparison of the nature of stacking fault predicted by the proposed models with experimental analysis through advanced transmission electron microscopy methods such as phase contrast HRTEM, aberration corrected Z-contrast HRSTEM-HAADF/ADF/BF, and iDPC imaging can provide further insights. In both the [2-1-10] and [0001] zone axes, high resolution phase contrast images reveal a variation in the array of bright columns at the faulted plane (Figures 1 and 3a). But under different defocus/thickness conditions, the contrast reverses, turning a bright column dark and vice versa. As such, it is challenging to pinpoint the exact deviation at the plane of fault by phase contrast imaging. Nevertheless, as Figure 3a illustrates, certain important material fingerprints have been deduced from HRTEM phase contrast images. First, the most advantageous direction for tungsten carbide growth is along the fault plane. Out of three sets, the set of {1100} planes with a greater number of stacking faults affects the crystal's final shape and growth direction. Second, a fault might cause the crystallite's thickness to vary by two to three atomic columns. Our third observation is that WC primarily displays faceted edges with [1100] and [0001] planes. At every step in the stacking defect, facet discontinuity is apparent.

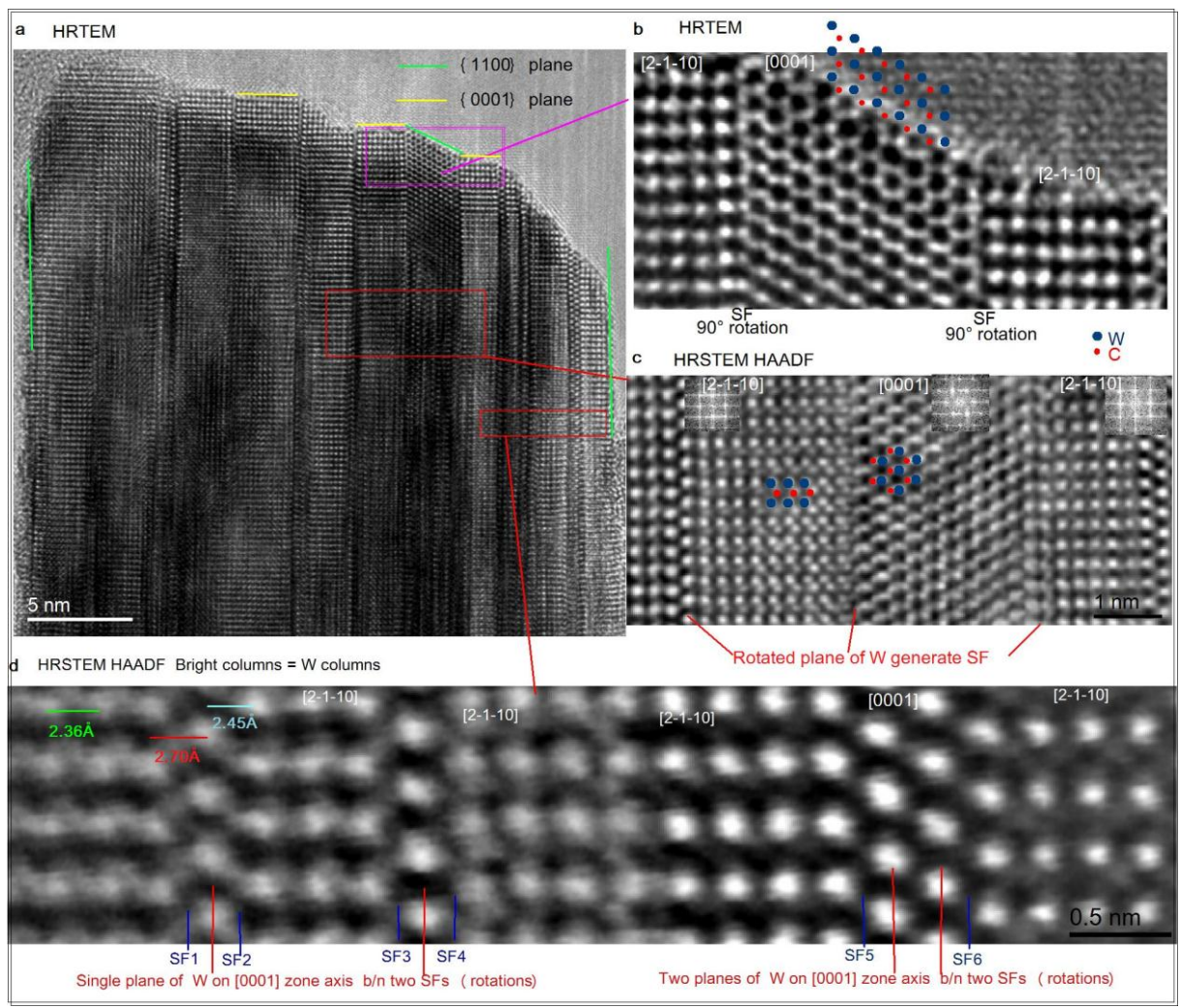

*Figure 3. The stacking fault generation by 90° rotation in the [1100] rotation axis is depicted in the (a-b) HRTEM and (c-d) HRSTEM images. (d) There is just one plane that is oriented at [0001] between SF1 and SF2. At SF2, the orientation returns to [2110]. Two rotated planes occur along the [0001] zone axis, between SF5 and SF6..*

The most significant conclusion comes from a deeper crystallographic examination of HRTEM, aside from the direct observations mentioned in the previous paragraph. At SF, {1-100} planes exhibit a 90° rotation in crystal orientation from [2-1-10] to [0001] as seen in the high-resolution image (Figure 3b-d). The rotation axis is [1100] zone axis. HRTEM image in Figure 3b illustrates how the column contrast varies on either side of the fault: on one side, the column contrast takes the form of a hexagonal honeycomb matrix, whereas on the other, bright columns organize with two-fold rectangular symmetry. The results are significantly straightforward when derived from Z-contrast aberration corrected STEM imaging (Figure 3c-d). The power spectra demonstrate that the crystal planes on both sides of the fault rotates. Carbon and tungsten columns projected as a hexagonal array in [0001], while in [2-1-10] orientation, atoms arrangement is rectangular. Figure 3d illustrates how frequently faults arise using STEM imaging in the HAADF (High angle annular dark field) mode, which scanned the location of W atoms. However, as can be observed in SF1:SF2 and SF3:SF4, the most frequent fault follows a single rotated plane.

There are now two inquiries that require responses: Initially, what is the source of the rotation, and secondly, does the rotation involve any slip? In that instance, what kind of thing is it?

Using iDPC imaging (Figure 4a-b), which can image high Z and low Z elements together and so captures the tungsten and carbon atoms together, the true cause of these rotations is attempted. Imaging the carbon plane between the faulted (rotated) tungsten planes was of particular interest. Focusing all three regions (faulted plane and atom column on either side) was a notable experimental challenge, due to defocusing issues caused by thickness variations associated with occurrence of the faults (faceted step formation and change in crystallographic orientation at fault as seen in Figure 3a). The tungsten and carbon column contrasts are depicted in Figure 4a's iDPC image; however, the atomic column contrast has disappeared on the fault plane. Using defocus condition fine-tuning, the atomic column contrast of the defective plane has been imaged (Figure 4b), where the matrix's column contrast on both sides decreases. As shown in Figure 4b, the iDPC images provide experimental evidence of double layer carbon planes between 2.82 Å dilated tungsten planes, which indicates that the WC structure's empty interstitials have been filled. The double layer of {1100} carbon planes rotates the adjacent {1100} tungsten plane by 90°, creating a [0001] zone axis. At the subsequent surplus carbon layers, the process is repeated and the planes spin back to the [2-1-10] direction (90°). The successive extra carbon layers may occur following a single rotated W plane (SF1/SF2 and SF3/SF4 in Figure 3d) or they may occur following many W planes (Figure 3c shows the rotation occurring after 11 W planes).

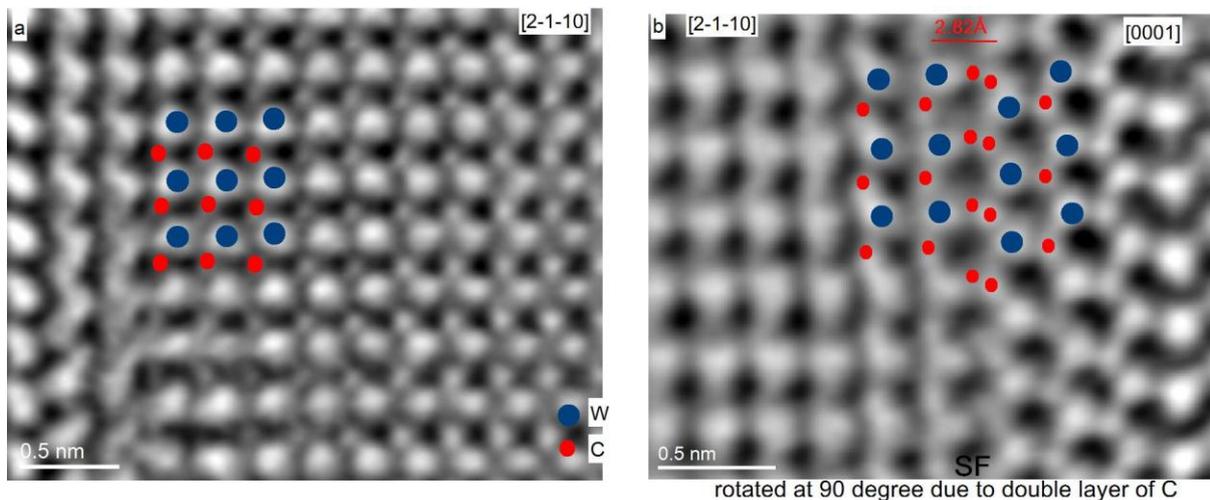

*Figure 4. iDPC images capture the tungsten and carbon atoms together: (a) Atomic columns of W and C in the matrix region. (b) Double layer carbon at the faulted plane*

The sequence of subsequent rotations/SFs affects interplanar distance at the fault. Figure 3d depicts a single plane (rotated at [0001]) with two distinct spacings for each, 2.70 Å and 2.45 Å, respectively, between SF1 and SF2. The interplanar spacing on fault SF5 is 2.62 Å, where two planes are rotated at the [0001] zone axis. A maximum interplanar distance of 2.82 Å is shown in figure 4b, where single SF is there between several periodic planes oriented at [0001] on the right and [2110] on the left. If the stacking faults are adjacent to each other, then the intrinsic arrangement of the atoms at fault causes the interplanar distance to vary on both sides as seen in SF1 and SF2. Another significant feature is the slide of tungsten plane adjacent to the SF. It is noted that following each SF, the tungsten atom's position in the {1100} plane changes. The movement may be inclined at any angle within a 360° range, even if the image appears to show a vertical slip. The rate of shift is dependent on the separation between the planes and varies at each SF. Based on direct experimental evidence, the following properties occur at the SF plane:

- Extra carbon exclusively found in the {1100} plane's empty pentahedral interstitials.
- Rearranged geometry of two-layer carbon plane causes {1100} plane to rotate at 90° angle.

- At the fault, the interplanar spacing, $d_{1100}$, has a discontinuity in its periodicity. The deviation, $\Delta d$ depends on the presence of subsequent stacking fault.
- Rotated tungsten plane exhibits slip, and the deviation of interplanar spacing affects the direction and location of the slip.

The dilatation of the interplanar distance between {1100} planes and the transition from hexagonality to a rectangular array of atomic packing are shown by theoretical studies of double-filled carbon layers (Models II and V: matching models to the experimental data). Because of the extra carbon planes, the structure near the defect is relaxed to a reorganized crystal with stretched planes (Figure 2c). For every atom in the relaxed crystal, the six fold symmetry changes to two-fold symmetry in the [0001] direction. However, in the experimental results, in order to maintain the crystal symmetry, the system spins at the faulty plane and obtains the rectangular pattern. Excellent agreement is seen between the theoretical calculation and the experimental data about the trend of variation of atomic columns' symmetry from hexagonal to rectangular at the fault. From theoretical data, another component that might be inferred is the likelihood of tungsten plane rearrangement within the fault. Rearranging at the defective plane can cause a change in the {1100} plane's rectangular packing.

**Carbon occupied in preferred pentahedra hence fault is only in {1100}! – Why?**

It appears that extra carbon atom layers in WC are only found in specific preferred orientations, limiting the defect's occurrence to {1-100}. This prompts even another query. The experimental findings clearly show that there is no random carbon occupancy and that the stacking fault's crystallographic preference in {1100}$_{WC}$ planes is consistent. An attempt is being made to understand the rationale behind the crystallographic selection of stacking faults by choosing {1100} plane for carbon filling.

The preference for defect may be significantly impacted by bonding behavior and electronic structure. Literature has extensively researched the W-W metal bond in WC with the hybridization of delocalized W5d orbitals [19]. In hexagonal WC, strong metal bonds tend to occur along the c-axis. Strong metal-metal bonding, demonstrated by the σ-σ bond, suggests that the tungsten atoms are metallically bonded [20]. In the present study, the relaxed WC structure models show overlapping Kahn sham orbitals only along the c-axis (Figure 5a). Qualitative assumption could suggest that the $dz^2$ degenerate state of the W atom's 5d orbital interacts along the c-axis. In the case of the

WC crystal structure, bonding along the inter-nuclear axis of metal atoms by σ-σ bond reduces the bond length along the c-axis and results in a c/a ratio smaller than one. For the same reason, unlike other transition metal mono-carbides, WC relaxes to a simple hexagonal structure rather than a NaCl-type structure [21].

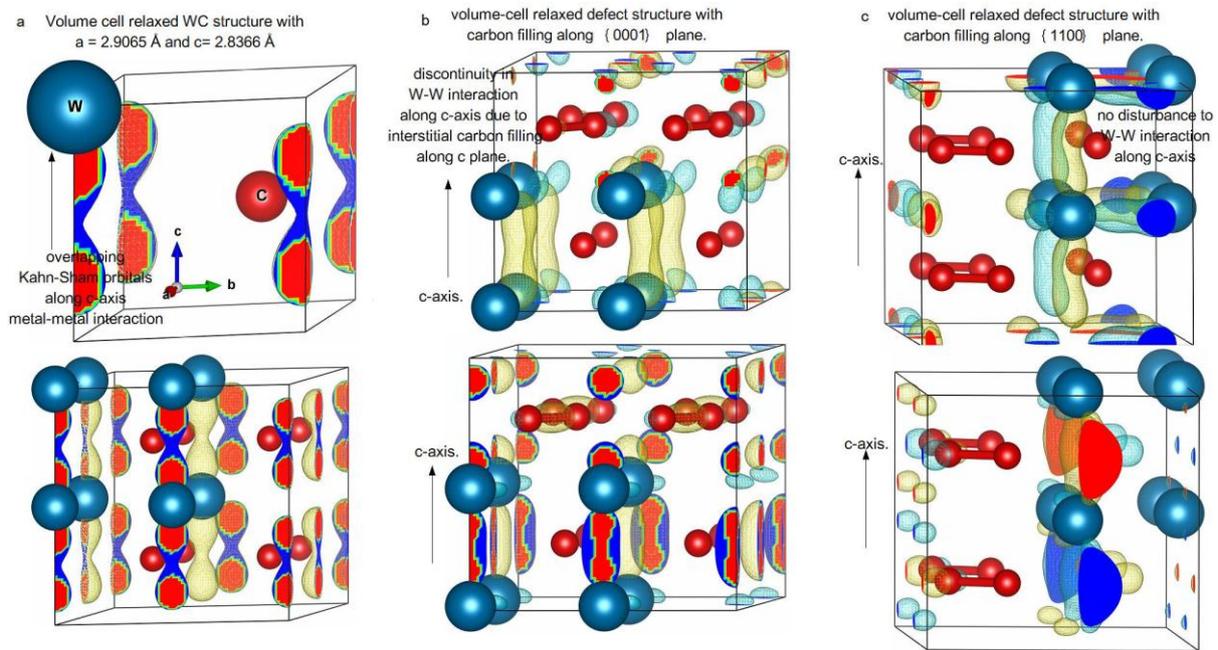

*Figure 5. 3D depiction of overlapping Kahn-Sham orbitals of volume cell relaxed WC and defected WC structures. (a) original WC crystal and 2x2x2 super cell, (b) defect introduced with carbon filling at interstitials along basal plane, {0001}, which breaks the strong metal - metal (W5d) interaction along c axis, and (c) interstitial filling of carbon along {1100} plane (as seen in experimental results), where the W-W strong d orbital interaction along the c axis is sustained. The carbon interstitial filling only occurs along {1100} planes due to the strong d-orbital interaction of tungsten atoms along the c axis; hence, the stacking defect is preferred along {1100} plane. The strength of the metal-metal bonding in WC, which adds to the materials' overall hardness, is also demonstrated by the results.*

More modelling research has been done to better understand the possibilities for random interstitial filling of carbon, using the introduction of a defect in the basal plane. The basal plane defect prevents Kahn Sham orbitals that overlap along the c-axis in the relaxed original structure (Figure 5b). Additionally, the dilated lattice spacing in (0001) planes contradicts the findings of the experiment. On the other hand, the energetically relaxed structure maintains the overlapping orbitals between tungsten atoms along the c-axis if faulted planes are introduced along {1100} (Figure 5c). The strong σ-σ metal-metal interactions are unaffected by faults along {1100}, according to the qualitative analysis. If the filled carbon interstitials are along {1100}, the dilation actually occurs along the {1100} plane in both the simulation and the experiment. The study shows, carbon cannot be arbitrarily filled in the pentahedra due to the strong W-W bonding along the c-axis. Since they are unlikely to interfere with metal-metal bonding, stacking faults in WC only arise along the {1100} planes.

Conclusion

To summarise, the atomistic cause of the stacking defect in the non-close packed system of the WC simple hexagonal is demonstrated by means of sophisticated atomic resolution methods, like aberration corrected microscopy combined with iDPC imaging and STEM/HAADF. Complementary crystallographic defects are modelled and theoretically simulated using DFT calculations to understand the experimental data. The WC crystal structure's pentahedral interstitials with additional carbon filling implement planar rearrangements as a rotating plane with glide that eventually manifest as the stacking defect. The favoured filling of carbon and the resulting stacking defect in the {1-100} planes of WC are caused by the strong metal-metal connection along the c-axis. These results demonstrate how stacking faults are introduced in non-close packed systems to accommodate carbon stoichiometric fluctuations in WC. The paper elucidates the contribution of interstitial elements to crystal faults, and highlights the significance of considering the involvement of low Z elements in defects.

Materials and methods

High density WC pellets were created using the Spark Plasma Sintering (SPS) once the temperature and uniaxial compaction pressure were optimised. Using a motor and pestle, sintered materials were ground into powder to create the TEM specimen. After suspending the powders in ethanol and dropping a few drops of solution onto the grid, fine particulates are gathered in a lacey carbon TEM grid. Thermofisher Scientific Talos TEM operating at 200 kV was used to perform selected area electron diffraction, X-ray energy dispersive spectroscopy (XEDS), and phase contrast TEM imaging. Thermofisher Scientific's Themis Z 60-300 ultra-high resolution analytical TEM, operated at 300 kV in aberration-corrected mode, has been used to perform atomic resolution STEM based analysis, such as STEM-HAADF, STEM-ABF, and (i/d)DPC.

The Quantum Espresso software package uses the Plane-Wave self-consistent field (PWscf) approach, which is implemented in a framework employing plane wave basis sets and pseudo-potentials, for DFT computations [22, 23]. Perdew, Burke, and

Ernzerhof (PBEsol), the exchange-correlation functional within the generalised gradient approximation (GGA) is used [24]. Energy cut-off and k-points are used by assuring convergence. For super cells up to 5x3x3, several iterations have been performed to determine the relaxed defect structure geometry of WC. Volume-cell relaxation calculations have been used for all defect structure super cells.